# Multiwavelength observations of the Blazar 4C + 28.07

Davit Zargaryan[1,2]★, Jonathan Mackey[1,2]★, Thibault Barnouin[3] and Felix Aharonian[1,2,4]

[1]*Dublin Institute for Advanced Studies, 31 Fitzwilliam Place, Dublin 2, Ireland*
[2]*Centre for AstroParticle Physics and Astrophysics, DIAS Dunsink Observatory, Dunsink Lane, Dublin 15, Ireland*
[3]*Université de Strasbourg, CNRS, Observatoire astronomique de Strasbourg, UMR 7550, Strasbourg F-67000, France*
[4]*Max-Planck-Institut für Kernphysik, PO Box 103980, Heidelberg D-69029, Germany*

**ABSTRACT**
The active galactic nucleus 4C + 28.07 is a flat-spectrum radio quasar, one of the brightest at $\gamma$-ray energies. We study its multiwavelength emission by analysing ∼12.3 yr of *Fermi-LAT* data in the $\gamma$-ray band and *Swift-X-Ray Telescope (XRT)/Ultraviolet/Optical Telescope (UVOT)* available data in X-ray and optical-to-ultraviolet bands. In the $\gamma$-ray band, five flaring periods have been detected, and quasi-simultaneously with these flaring times, the X-ray and *UVOT* data detected by *Swift-XRT/UVOT* have also been analysed. In one of the brightest flare periods (Flare 5; observed on 2018 October 12), the $\gamma$-ray flux reached $(6.7 \pm 0.81) \times 10^{-6}$ photon cm$^{-2}$ s$^{-1}$ (∼31 × higher than the mean flux over 12.3 yr) with detection significance of $\sigma = 6.1$. The estimated variability time (∼2 h) constrains the $\gamma$-ray emitting region size to $\leq 9 \times 10^{14}$ cm, which is close to the black hole radius. The spectral energy distributions (SEDs) in the $\gamma$-ray band for the ∼12.3 yr of data show an early cut-off at ∼14 GeV; beyond ∼60 GeV, however, the spectrum hardens and is detected up to ∼316 GeV. Similar spectral behaviour is also noticeable for the SEDs of flares, which can be linked to the photon absorption by the emitting region's internal and external narrow-band radiation fields. In the quiescent period, the $\gamma$-ray emission was described by the synchrotron self-Compton scenario, while the external photons contributions from the disc and the broad-line region were required to explain the short-term flaring $\gamma$-ray emission. Considering the significance of the obtained results from 4C + 28.07, we compared the parameters with 3C 279 and M87, to motivate further studies.

## 1 INTRODUCTION

Blazars are a subclass of active galactic nuclei (AGNs), considered one of the most violent places in the Universe through observations across the whole electromagnetic spectrum, from radio to high energy (or very high energy) [HE; >100 MeV (VHE; >100 GeV)] $\gamma$-ray bands. They are characterized by jets, where highly relativistic motion of the emitting region is directed toward our line of sight, and the particles can continuously accelerate within them (Begelman, Blandford & Rees 1984; Urry & Padovani 1995). Based on the features of spectral emission lines, generally, blazars are divided into two primary classes: i) flat spectrum radio quasars (FSRQs) characterized by prominent emission lines (equivalent width >5 Å) and ii) BL Lacertae which have weak or undetected spectral lines (Stickel et al. 1991; Urry & Padovani 1995).

The broad-band spectral energy distribution (SED) of blazars is characterized by two peaks. The first, low-energy peak which extends from radio to ultraviolet (UV)/X-ray band and can be explained by synchrotron emission of relativistic electrons accelerated in a magnetized supersonic jet (Blandford & Rees 1978; Landau et al. 1986; Ghisellini, George & Done 1989). The second component of the SED dominates from X-ray to high energies (HE; >100 MeV), and its origin is a matter of debate. For instance, in the scope of leptonic scenario the HE/VHE component is explained by inverse Compton (IC) scattering of synchrotron photons [so-called synchrotron self-Compton (SSC)] (Ghisellini, Maraschi & Treves 1985; Bloom & Marscher 1996) or of the photons with external origin [external inverse Compton (EIC)] (Sikora, Begelman & Rees 1994; Błażejowski et al. 2000; Böttcher 2007; Ghisellini & Tavecchio 2009). The external radiation field contributes additional seed photons to make the IC mechanism a convincing explanation for the HE/VHE $\gamma$-ray emission. In this mechanism, depending on the emitting region's distance from the supermassive black hole (SMBH), the energy density of the external photon fields may vary. Generally, in the scope of blazars studies, the external photon field originates in the accretion disc, broad-line region (BLR), and dusty torus (Dermer & Schlickeiser 1993; Sikora et al. 1994; Błażejowski et al. 2000; Zargaryan et al. 2017; Costamante et al. 2018). Additionally, studies have proposed an alternative hadronic mechanism, i.e. the synchrotron emission of relativistic protons, to explain the second peak of broad-band SED (Aharonian 2000; Mücke & Protheroe 2001; Zargaryan, Romoli & Aharonian 2019).

Blazars are recognized as powerful accelerators (Aharonian 2000; Aharonian et al. 2002; Lemoine & Waxman 2009; Abdalla et al. 2020) with time-variable emission that is quite evident in the HE γ-ray band. Several flaring and variable γ-ray AGNs (e.g. 3C 454.3, 3C 279, PKS 2155-304, etc.) have variable energy flux at a time-scale of a few minutes (Aharonian et al. 2007; Striani et al. 2010; Hayashida et al. 2015) with amplitude at the $10^{-10}$ erg cm$^{-2}$ s$^{-1}$ level. The energies of the parent electron/proton population of these extreme fluxes are evaluated to be in TeV/PeV bands, accelerated in the relativistic jets and the study of these sources are useful for investigating a range of physical phenomena that take place in AGNs.

★ E-mail: dzargaryan@cp.dias.ie (DZ); jmackey@cp.dias.ie (JM)

The blazar 4C + 28.07 is one of the brightest FSRQs, located at redshift $z = 1.21$ and with SMBH mass $M_{BH} = (1.65^{+1.66}_{-0.82}) \times 10^9$ M$_\odot$ (Shaw et al. 2012). In the radio band, very long baseline interferometry observations show a parsec-scale, one-sided jet structure and, based on the moving features in the jet, the apparent median jet speed of $(10.11 \pm 0.39)c$ is estimated (Lister et al. 2009, 2019). The jet has a rich, bright structure from the Very Large Array image, and it is extending towards the north direction and eventually bends sharply. In the X-ray band, *Chandra* X-ray observations of 4C + 28.07 show extended and knotty jet morphology, with the jet bending also noticeable (Marshall et al. 2011).

In the γ-ray band, the source has been detected by *Fermi-LAT* above 100 MeV and is included in the 4FGL (Abdollahi et al. 2020), 3FGL (Ackermann et al. 2015), and 2FGL (Nolan et al. 2012) *Fermi* catalogues. Also, this source is included in the second *Fermi* flares catalogue (Abdollahi et al. 2017). Most recently, Das, Prince & Gupta (2021) presented the detailed analysis of ~12 yr of *Fermi-LAT* data of 4C +28.07. Analysing the light curve of 4C + 28.07 in 10-d binning, they found three distinctive flaring states and these flares have been further analysed in 3-d time bin. Also, they have investigated the correlation between radio and γ-ray emission during flaring times. They explained γ-ray production can originate near the BLR region and efficiently explained by EIC mechanism.

In this work, we present detailed multiwavelength studies of 4C + 28.07 FSRQ by analysing *Fermi-LAT* and *Swift* archival data. In Section 2, we present the observational data and data-reduction methods. Section 3 shows the obtained results. Then, in Section 4, we present the theoretical modelling of broad-band SED. Finally, Sections 4 and 5 are the discussion and conclusion, respectively, of this work.

## 2 OBSERVATIONS AND DATA ANALYSIS

### 2.1 Fermi-LAT

*Fermi-LAT* is a pair-conversion space-based telescope designed to detect HE γ-rays in the energy range of 20 MeV to 1 TeV (Atwood et al. 2009). Since 2008 August 4, this telescope is in operation by scanning the entire sky in 3 h, and the continuous monitoring of the γ-ray sky has made it possible to detect thousands of γ-ray sources. In this work, we study ~12.3 yr of *Fermi-LAT* Pass 8 data, from 2008 August 8 to 2020 September 19 (MET 239557417 - 622166405), towards the sky position of 4C + 28.07 FSRQ (RA, Dec) = (39.474, 28.805). For further analysis, we defined the region of interest (ROI), 15° circular region around the target position, in the energy range of 100 MeV to 1 TeV. The latest version of FERMIPY (version 1.0.0; Wood et al. 2017) and FERMITOOLS v2.0.0[1] software packages have been used for analysing γ-ray data with the latest instrument response function P8R3_SOURCE_V3. To exclude the contamination of γ-rays, occurring from interaction with the atmosphere, a zenith angle cut 90° has applied. We build a model file, which includes all sources in the ROI, based on the latest *Fermi-LAT* 4FGL catalogue (Abdollahi et al. 2020). During the analysis, all source parameters inside the ROI were left as free parameters, while sources parameters beyond the ROI were fixed, according to the 4FGL catalogue. To take into account the diffuse background emission, the Galactic diffuse emission model gll_iem_v07.fits and an extragalactic isotropic emission component iso_P8R3_SOURCE_V2_v1.txt are considered by adding to the model file during the fitting. The detection significance for each source in ROI has been evaluated by applying the Test Statistics (TS) defined as TS = $2\Delta$log(likelihood) between models with and without the source (Mattox et al. 1996).

### 2.2 Swift-XRT/UVOT

*Swift* (Gehrels et al. 2004) is a space-based telescope carrying instruments including the *X-Ray Telescope (XRT)*, *Ultraviolet/Optical Telescope (UVOT)*, and a hard *XRT* (*Burst Alert Telescope*) opening opportunities to investigate the sky in X-ray and optical/UV bands simultaneously with other wavelengths. In this paper, we analysed available *Swift* data for 4C + 28.07 in the context of multiwavelength observations, with HE γ-rays. The *Swift*-XRT (Burrows et al. 2005) data were processed using *XRT* Data Analysis Software package (XRTDAS; v. 3.4) following the standard procedure based on the most recent calibration data base. All observations were made in photon-counting mode, since the detected fluxes were at a low level (0.1 mCrab). For the extraction of the spectrum, a circular region with a radius of 30 pixels (~71 arcsec) and an annulus with its inner and outer radii between 80 (~190 arcsec) and 120 pixels (~280 arcsec) around of the source have been defined as source and background regions, correspondingly. We did not find any of the spectrum affected by pile-up, since the counts rate for all observations were below 0.5 count s$^{-1}$. For the fitting of the spectrum, we used XSPEC v12.11.1 X-ray spectral fitting package.[2] The spectra were fitted by using $\chi^2$-minimization technique assuming rebinned at least 20 counts per bin, in the energy range of 0.3–10 keV. Table 1 lists all of the observations that were analysed in this work. 1.5 + the spectra for all observations are well fitted by an absorbed power-law (PL; wabs∗powerlaw) model with column density $N_H = 0.756 \times 10^{21}$ cm$^{-2}$ from the Survey of Galactic H I data base (Kalberla et al. 2005). The best-fitting results and derived values are reported in Table 1. The X-ray spectra for all observations were well fitted by a PL with a mean value of photon index ($\Gamma_X \sim 1.38$).

The *Swift-UVOT* telescope (Roming et al. 2005) observed 4C + 28.07 FSRQ simultaneous with the *Swift-XRT*. The *UVOT* observations have been taken in different filters (V: 5440 Å, B : 4390 Å, U : 3450 Å, W1 : 2510 Å, M2 : 2170 Å, and W2 : 1880 Å) and each one analysed separately. Using standard 5 arcsec as a source region and 27 arcsec (inner) and 35 arcsec (outer) radii as a background region around the source, the photometry were computed for all observations. The magnitudes were computed using the uvotsource tool (HEASOFT v6.21), corrected for extinction according to Roming et al. (2009) using the reddening coefficient $E(B - V)$ from Breeveld et al. (2011) and Schlafly & Finkbeiner (2011). Afterwards, we converted counts rates to fluxes following and using calibration factors from Poole et al. (2008). The observational data and analysis results are presented in Table 2.

---

[1] https://fermi.gsfc.nasa.gov/ssc/data/analysis/documentation/

[2] https://heasarc.gsfc.nasa.gov/xanadu/xspec/

**Table 1.** The spectral analysis results of *Swift*-XRT observations. Column 3 is the exposure time, Column 4 is the logarithm of X-ray flux in the 0.3−10 keV energy band in units of erg cm$^{-2}$ s$^{-1}$, and column 5 is the fitted PL spectral index.

| | | *Swift*-XRT | | | |
|---|---|---|---|---|---|
| Obs. ID | Time (MJD) | Exposure time (s) | $\log_{10}(F_X)$ | $\Gamma_X$ | Reduced $\chi^2$(dof) |
| 00036189003 | 54715 | 2397 | $-11.65 \pm 0.06$ | $1.50 \pm 0.18$ | 1.05(74) |
| 00036189004 | 55232 | 1231 | $-11.44 \pm 0.11$ | $0.98 \pm 0.25$ | 1.12(41) |
| 00036189005 | 55232 | 2525 | $-11.95 \pm 0.07$ | $1.57 \pm 0.21$ | 0.94(50) |
| 00036189006 | 55840 | 4922 | $-11.31 \pm 0.03$ | $1.36 \pm 0.08$ | 1.06(224) |
| 00036189007 | 55843 | 2879 | $-11.42 \pm 0.03$ | $1.57 \pm 0.11$ | 0.82(156) |
| 00036189008 | 55961 | 5107 | $-11.38 \pm 0.03$ | $1.42 \pm 0.08$ | 1.09(242) |
| 00036189010 | 57989 | 3219 | $-11.34 \pm 0.04$ | $1.38 \pm 0.12$ | 0.94(134) |
| 00036189011 | 57992 | 174.8 | $-11.49 \pm 0.20$ | $1.47 \pm 0.27$ | 0.60(17) |
| 00036189012 | 58378 | 2977 | $-11.29 \pm 0.03$ | $1.40 \pm 0.10$ | 0.77(187) |
| 00036189013 | 58384 | 2859 | $-11.36 \pm 0.04$ | $1.20 \pm 0.12$ | 0.84(125) |

**Table 2.** Fluxes measured from the *Swift* optical and UV data at the time simultaneously with *Swift* X-ray.

| | | Optical *Swift*-UVOT fluxes ($\times 10^{-12}$ erg cm$^{-2}$ s$^{-1}$) | | | | | |
|---|---|---|---|---|---|---|---|
| Obs. ID | Time (MJD) | V | B | U | W1 | M2 | W2 |
| 00036189003 | 54715 | – | – | – | $0.78 \pm 0.03$ | – | – |
| 00036189004 | 55232 | $1.73 \pm 0.23$ | $1.46 \pm 0.21$ | $1.29 \pm 0.13$ | $0.81 \pm 0.06$ | – | $0.69 \pm 0.06$ |
| 00036189005 | 55232 | – | – | $1.48 \pm 0.05$ | – | – | – |
| 00036189006 | 55840 | – | – | $2.52 \pm 0.07$ | – | – | – |
| 00036189007 | 55843 | – | – | $2.34 \pm 0.06$ | – | – | – |
| 00036189008 | 55961 | $3.05 \pm 0.11$ | $2.42 \pm 0.09$ | $2.14 \pm 0.06$ | – | – | – |
| 00036189010 | 57989 | $9.28 \pm 0.34$ | $7.53 \pm 0.21$ | $6.71 \pm 0.18$ | $4.67 \pm 0.17$ | $3.56 \pm 0.16$ | $3.31 \pm 0.12$ |
| 00036189011 | 57992 | $6.85 \pm 1.12$ | $6.38 \pm 0.87$ | $5.33 \pm 0.73$ | $4.26 \pm 0.67$ | $3.35 \pm 0.60$ | $2.48 \pm 0.42$ |
| 00036189012 | 58378 | $14.11 \pm 0.53$ | $12.26 \pm 0.34$ | $10.73 \pm 0.35$ | $7.69 \pm 0.28$ | $6.02 \pm 0.17$ | $4.69 \pm 0.17$ |
| 00036189013 | 58384 | $9.9 \pm 0.37$ | $8.25 \pm 0.26$ | $6.98 \pm 0.26$ | $4.65 \pm 0.18$ | $3.66 \pm 0.18$ | $3.31 \pm 0.12$ |

## 3 RESULTS

### 3.1 $\gamma$-ray variability

One of characteristic features of blazars is the flux variability over time, which is significantly noticeable in the HE $\gamma$-ray band. The study of flux temporal variation from blazars could be key for answering several essential questions, e.g. the acceleration region size and emission mechanisms. The observed variability time $\tau$ could constrain the emission region size, based on $R \leq (\delta c \tau)/(1 + z)$ equation, where $\delta$ is Lorentz Doppler factor, c is the speed of light, and z is the redshift of the source. Considering that the jets in blazars are aligned close to the line of sight of the observer, the observed emission from the jet is Doppler boosted by factor $\delta = 1/\Gamma(1 - \beta\cos\theta)$, where $\Gamma$ is the bulk Lorentz factor, $\theta$ is the angle between the source jet-speed axis and the line of sight, and $\beta = v/c$ is responsible for the jet speed. Obviously, for this limitation of the emission region size, the Doppler factor plays a key role and requires proper estimation. Studies have shown that for Blazars, it is typically higher than 10 and in the case of powerful accelerators, the observed variability time $\tau$ could be <1 d. We can express the emission region size as

$$R \leq 2.5 \times 10^{16} \text{ cm} \left(\frac{\delta}{10}\right) \left(\frac{\tau}{1 \text{ day}}\right) \left(\frac{1}{1+z}\right). \quad (1)$$

The FSRQ 4C + 28.07 is one of the most variable AGNs in the HE $\gamma$-ray band, where several flaring periods have been investigated. By dividing the whole time range into weekly (7-d) time bins, we found that $\gamma$-ray flux shows flaring activities in more than five periods, where the flux level is about a factor of $\geq$5 times higher comparing to its quiescent level as shown in Fig. 1a. We defined an orange dashed line $f = 5 \times 10^{-7}$ photon cm$^{-2}$ s$^{-1}$ as the baseline level, above which the flaring periods are considered. Also these $\gamma$-ray flaring periods are found in Das et al. (2021). Based on the weekly light curve, five flaring periods (see Fig. 1a, numbered as 1–5) have been selected for more detailed investigation with smaller time bins ($\leq$1 d). At the first stage, daily light curves are generated for each of the flaring periods where $\gamma$-ray flux variations have been detected. Furthermore, for Flare 5, the Bayesian Blocks approach (Scargle et al. 2013) is applied to identify statistically significant flux variations in optimally spaced time intervals. For the Bayesian Block analysis, we used the `astropy version 4.2` Python package[3] assuming a false alarm probability $p_0 = 0.05$. During Flare 5, six blocks were identified in the daily light curve, and the shortest and most significant (>13$\sigma$) one is Block 4 (see Fig. 2, left-hand panel). Block 4 was further investigated in the 30-min light curve and the analysis detected three subblocks with significant variation, the shortest of which (subblock 2) has duration 2 h (Fig. 2, right-hand panel). This shows that 4C + 28.07 can be variable on time-scales of $\leq$2 h, with our sensitivity to shorter time-scales limited by photon statistics.

As a result, the period of strongest and most rapid $\gamma$-ray variability in Flare 5 was investigated. Notably, during the 30 min in Flare 5 period (observed on 2018 October 12), the $\gamma$-ray flux (above 100 MeV) reached $f = (6.7 \pm 0.81) \times 10^{-6}$ photon cm$^{-2}$ s$^{-1}$ with detection significance of $\sigma = 6.1$, which corresponds to apparent $\gamma$-ray luminosity of $L_{\text{app}} = 4\pi D_L^2 f = 3.6 \times 10^{49}$ erg s$^{-1}$, for a luminosity distance of $D_L = 8.38$ Gpc. This is about ∼100 times than the Eddington luminosity; however, it is assumed that the $\gamma$-ray radiation is produced in relativistic jet ($\delta \gg 1$), which is closely directed to the observer (Begelman et al. 1984; Urry & Padovani 1995). The intrinsic luminosity $L_{int} = L_{app}\delta^{-4} \approx 10^{45}$ erg s$^{-1}$, by assuming relativistic Doppler factor $\delta \geq 10$ typical for blazars (see also Section 4). In Fig. 3, the light curves of all five flaring periods, for data binned on 1-d and shorter time-scales, are depicted. For all flaring peak times, when detection statistics allowed (TS > 250), the temporal flux variations for small time-scales ($\leq 6$ h) were investigated and plotted with the daily light curves. To display the data more clearly, only flux bins $\geq 3\sigma$ are shown. Generally, the fractional variability parameter ($F_{var}$) for all of the light curves we investigated is $F_{var} > 0.6$ (Vaughan et al. 2003), showing clear variability in the $\gamma$-ray band.

---
[3] https://docs.astropy.org/en/stable/api/astropy.stats.bayesian_blocks.html

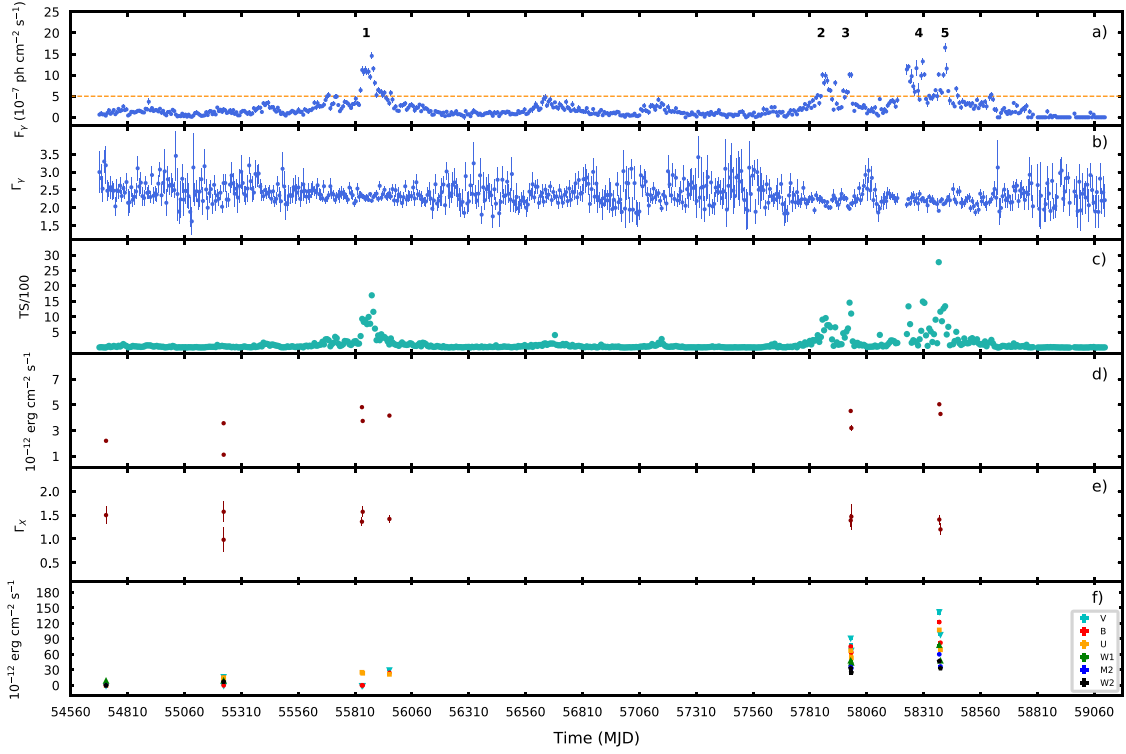

**Figure 1.** The multifrequency light curves of 4C + 28.07: a) $\gamma$-ray weekly (7-d bins) light curve, b) $\gamma$-ray spectral-index evolution for each time-bin, c) TS value for each bin, d) X-ray light curve, e) X-ray spectral-index, and f) *Swift-UVOT* flux for V, B, U, W1, M2, and W2 bands.

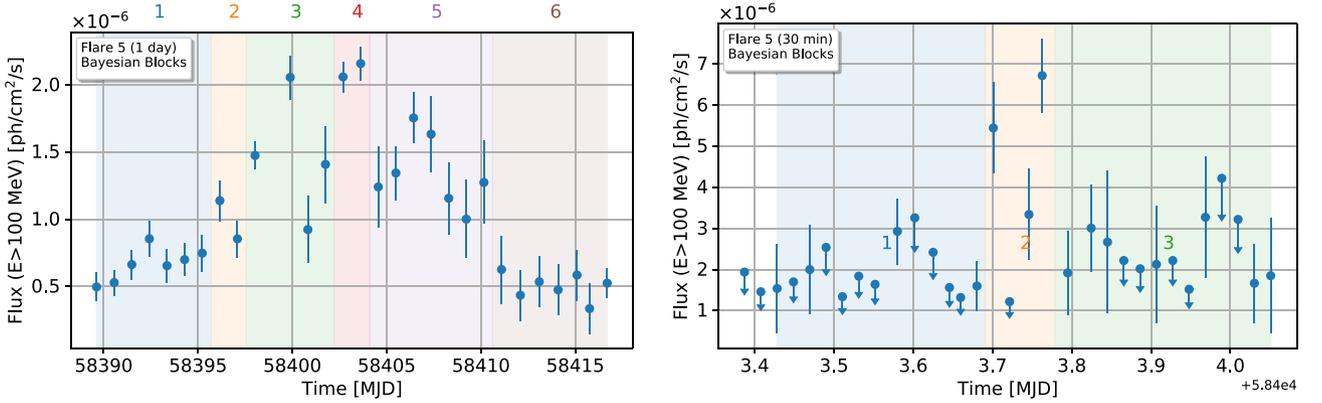

**Figure 2.** Flare 5 light curve for 1-d (left-hand panel) and 30-min (right-hand panel) bins. Bayesian analysis approach conducted for defining solid blocks and estimating variability time.

### 3.2 $\gamma$-ray spectral analysis

The $\gamma$-ray spectrum of 4C + 28.07 has been investigated from 100 MeV to 1 TeV. Based on the results of the $\gamma$-ray light curve, the spectra of both quiescent and flaring periods have been studied individually. For the quiescent period, we consider the whole ∼12.3-yr data since, on average, the flux level for the entire period is quite relaxed. For the flaring periods, we follow the time column mentioned in Table 3. Two spectral models were applied to describe the $\gamma$-ray spectrum. First, a PL

$$\frac{dN}{dE} = N_0 \left(\frac{E}{E_0}\right)^{-\alpha}, \quad (2)$$

where $N_0$ is the spectral normalization at the scale energy $E_0$ and $\alpha$ is the PL index. Secondly, we used the exponential cutoff power-law

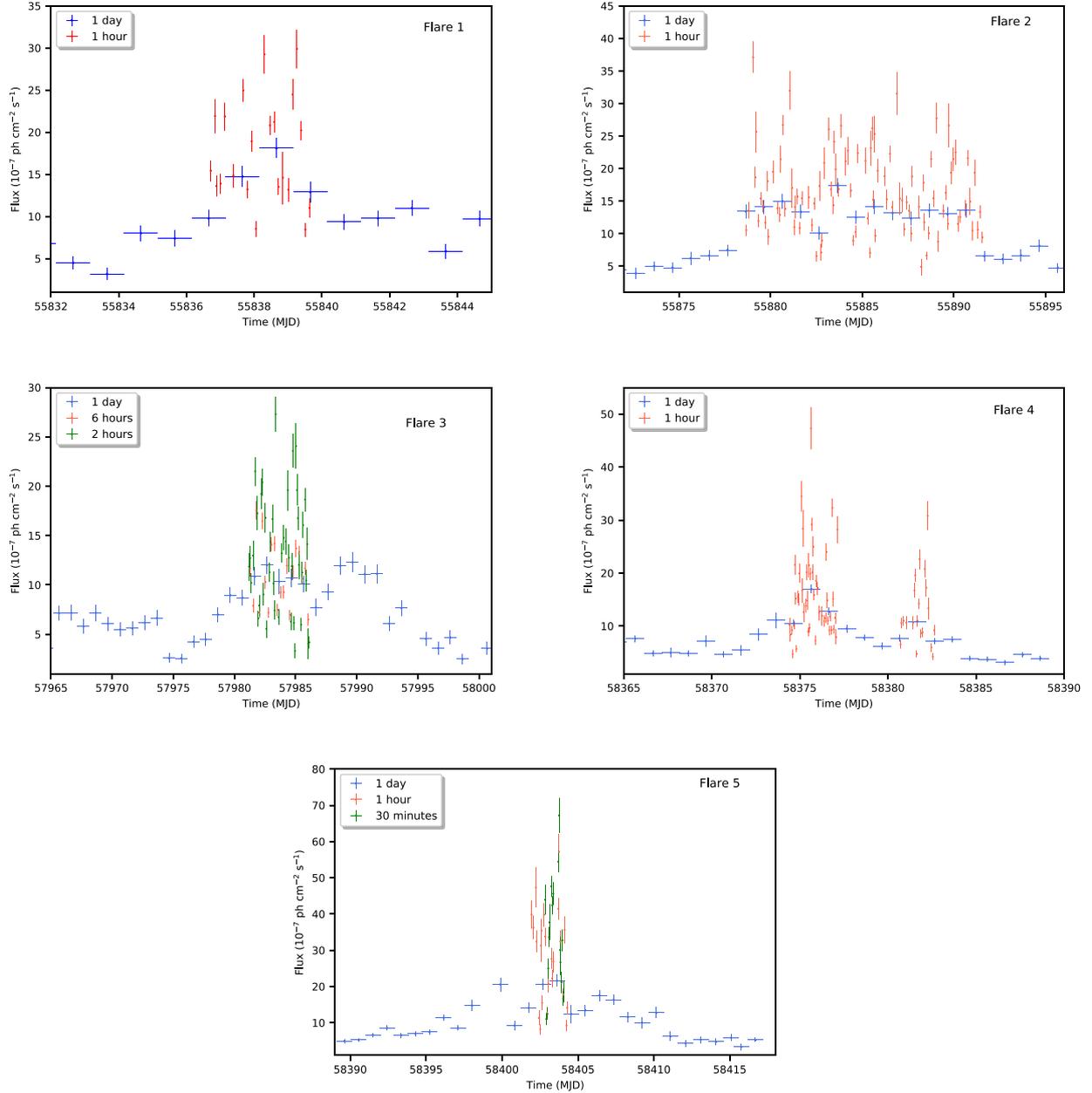

**Figure 3.** Light curves for the five identified flaring periods of 4C + 28.07: for all time periods as colour-coded in the legend of each panel.

**Table 3.** The spectral analysis results of the $\gamma$-ray data with corresponding 1$\sigma$ errors. Column 3 shows $\gamma$-ray flux in units $\times\,10^{-7}$ photon cm$^{-2}$ s$^{-1}$; column 4 the fitted PL spectral index for the PL model; and column 5–7 the fitting parameters for the ECPL model.

| Period | Time | Flux | Index (PL) | Index (ECPL) | Beta | Ecut (MeV) |
|---|---|---|---|---|---|---|
| Whole time | 2008 Aug 08–2020 Sep 19 | 2.08 ± 0.021 | 2.27 ± 0.04 | 2.179 ± 0.01 | 0.99 ± 0.08 | 14 294.0 ± 1369 |
| Flare 1 | 2011 Sep 28–2011 Oct 10 | 10.97 ± 0.52 | 2.33 ± 0.05 | 1.97 ± 0.226 | 1.056 ± 0.5 | 2214 ± 1722 |
| Flare 2 | 2011 Nov 08–2011 Nov 30 | 12.57 ± 0.41 | 2.27 ± 0.04 | 1.39 ± 0.08 | 0.27 ± 0.01 | 9.398 ± 3.89 |
| Flare 3 | 2017 Jul 31–2017 Sep 04 | 9.03 ± 0.36 | 2.04 ± 0.03 | 1.06 ± 0.02 | 0.22 ± 0.002 | 1.016 ± 0.13 |
| Flare 4 | 2018 Sep 04–2018 Sep 29 | 9.21 ± 0.34 | 2.01 ± 0.02 | 1.99 ± 0.03 | 0.004 12 ± 0.0001 | 1020 ± 253 |
| Flare 5 | 2018 Sep 29–2018 Oct 24 | 12.89 ± 0.06 | 2.11 ± 0.03 | 1.89 ± 0.13 | 1.02 ± 0.32 | 4873 ± 2729 |

(ECPL) model, i.e.

$$\frac{dN}{dE} = N_0 \left(\frac{E}{E_0}\right)^{-\alpha} \exp\left[-\left(\frac{E}{E_c}\right)^{\beta}\right], \qquad (3)$$

where $E_c$ is the cutoff energy and $\beta$ is the cutoff parameter. The spectral plots for each period are presented in Fig. 4.

The SEDs for the quiescent and flaring periods have been built according to the periods and time column mentioned in Table 3. The whole time $\sim$12.3-yr $\gamma$-ray flux above 100 MeV is $F_\gamma =$ $(2.02 \pm 0.021) \times 10^{-7}$ photon cm$^{-2}$ s$^{-1}$ with $\alpha = 2.27 \pm 0.04$ photon index and higher detection significance ($\sigma = 211$). The $\gamma$-ray spectrum shows complex structure by having cut-off at $\beta \sim 14$ GeV; however, it extends up to $\sim$316 GeV with $\sim$2.1$\sigma$ confidence. Interestingly, beyond $\sim$60 GeV, the spectrum dramatically hardens (Fig. 4, top-left panel). Also, the same spectral behaviour is noticeable for the flaring time periods. The $\gamma$-ray spectral analysis results are shown in Table 3 and depicted in Fig. 4.

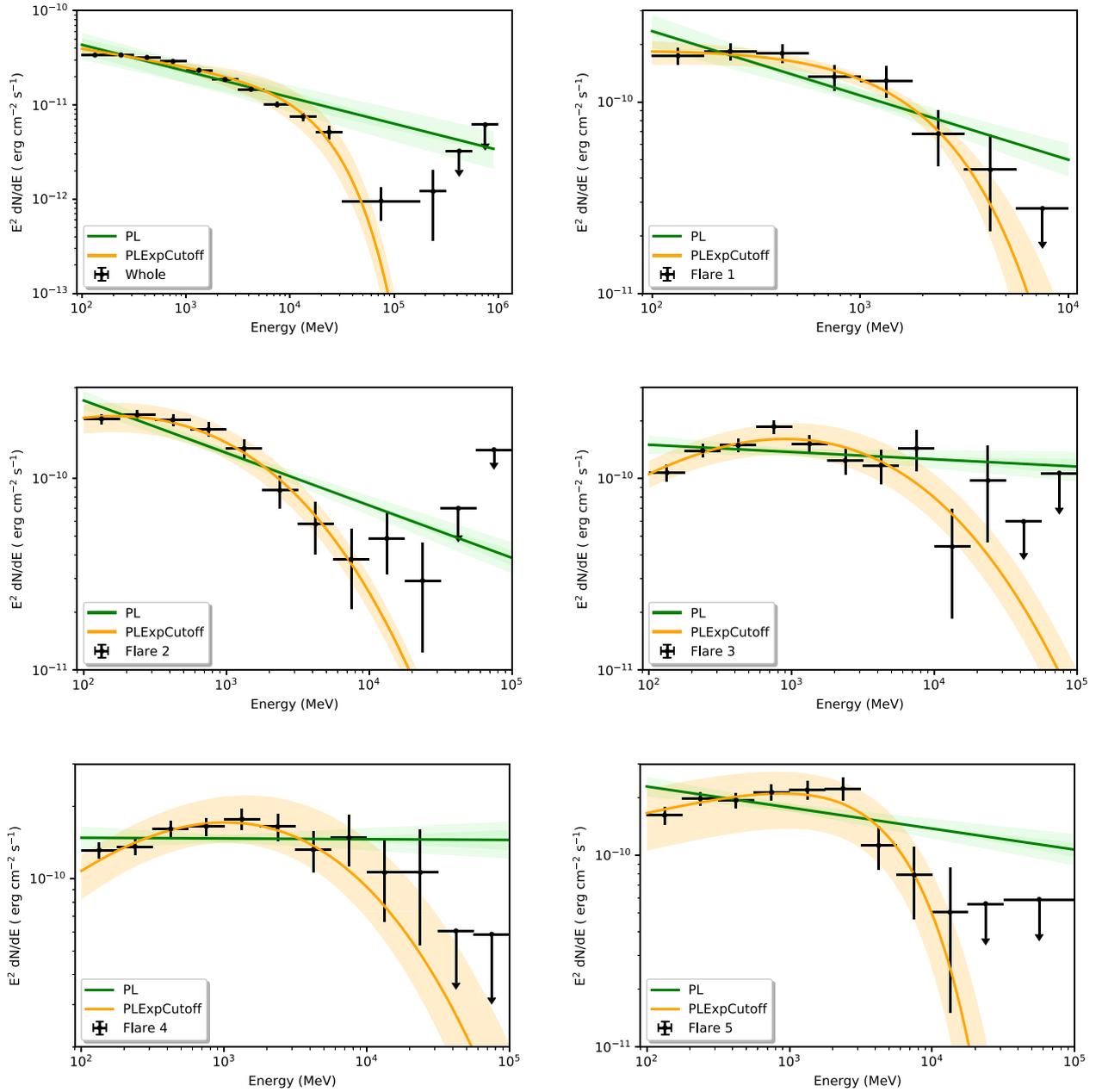

**Figure 4.** The SEDs for the full time range and five flaring periods are depicted, respectively. The data points are fitted with PL (green) and power law with exponential cutoff (PLExpCutoff; orange) functions.

## 4 BROAD-BAND SEDS AND THEORETICAL MODELLING

The broad-band multiwavelength SEDs of 4C + 28.07 have been built by analysing data in the $\gamma$-ray, X-ray, and optical/UV bands detected by *Swift* and *Fermi-LAT* instruments. Fig. 5 presents the multiwavelength SEDs of 4C + 28.07 in flaring and quiescent periods. The archival data from radio to near-infrared (grey colour) are taken from the ASI science data centre,[4] and the optical/UV/X-ray to $\gamma$-rays are quasi-simultaneously measured data. From Fig. 5, it is noticeable that during the flaring periods the level of the multiwavelength flux from UV/optical to $\gamma$-ray bands is comparatively higher than its average (quiescent; blue colour) level. This flux level difference across different wavebands can constrain the different radiating particles, emission mechanisms, or the different locations of the acceleration regions that are responsible for the radiation.

[4] https://tools.ssdc.asi.it/SED/

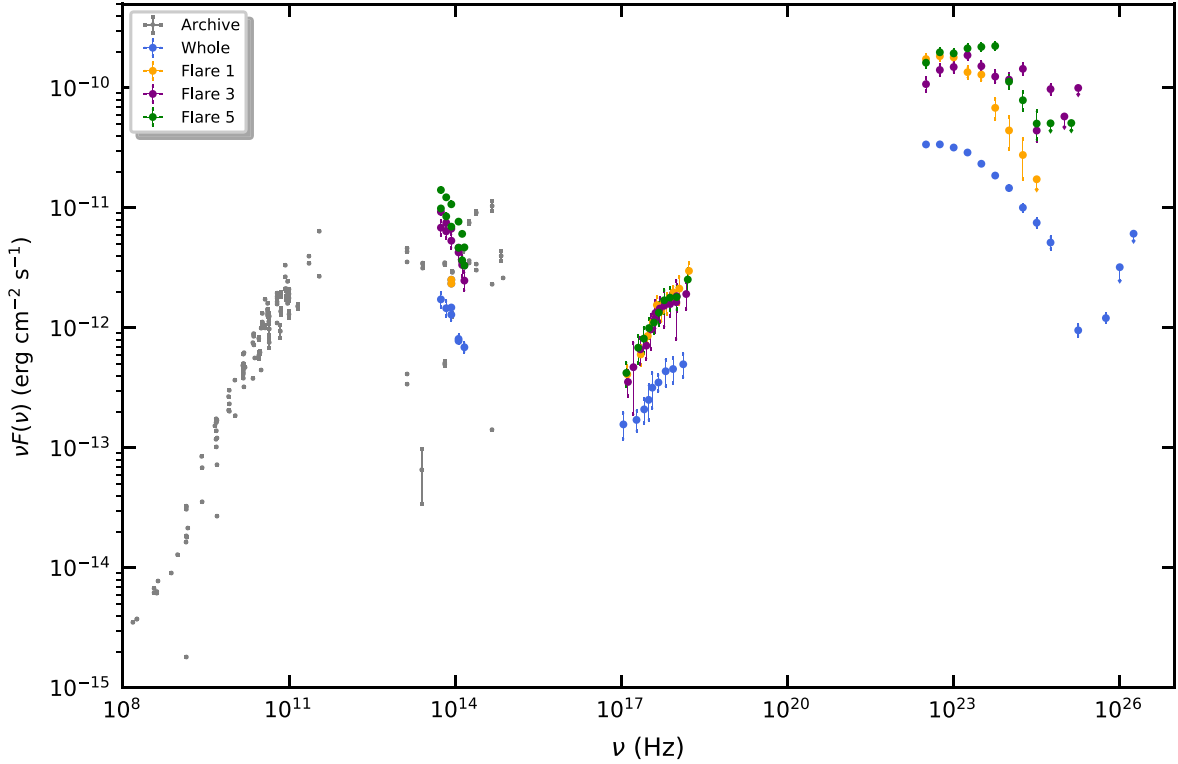

**Figure 5.** The multiwavelength SEDs for the whole 12.3-yr data set, and flare periods 1, 3, and 5 are presented.

For the further phenomenological explanation of the broad-band SEDs and the flaring behaviour of 4C + 28.07, theoretical modelling has been implemented by assuming different scenarios. Generally, it is assumed that the emitting region is a spherical blob with radius $R$, which moves along the jet with bulk Lorentz factor $\Gamma$, and the inclination angle between the source and the observer is $\Theta$. The blob is magnetized (magnetic field strength $B$), and the region is filled with non-thermal relativistic particles, with the following distribution

$$N(\gamma) \sim (\gamma)^{-\Gamma_e} \exp[-(\gamma/\gamma_c)], \gamma > \gamma_{\min}, \quad (4)$$

where $\gamma$ is the Lorentz factor of electrons, $\Gamma_e$ is the PL index, and $\gamma_c$ and $\gamma_{\min}$ describe the cut-off and the minimum energy of the electrons.

In the scope of modelling, it is assumed that the low-energy component of SED (from radio to optical/UV) is the result of synchrotron emission of relativistic electrons, while the HE component is believed to have SSC or EIC origin (Blandford & Rees 1978; Landau et al. 1986; Sikora et al. 1994; Bloom & Marscher 1996). The radio archival data is considered to be an upper limit and assumed to originate from different and extended regions.

Since the jet is highly relativistic ($\Gamma \gg 1$) and the external photons penetrate into the emission region under the angles ($\Theta > 1/\Gamma$), the transformation of the external fields to the emission region rest frame has been evaluated according to (Dermer & Schlickeiser 1993; Dermer 1995; Georganopoulos, Kirk & Mastichiadis 2001). For the modelling, we used the *jetset* fitting package (Massaro et al. 2006; Tramacere et al. 2009; Tramacere, Massaro & Taylor 2011; Tramacere 2020).

First, we modelled the broad-band SED for quiescent period as shown in Fig. 6 (left-hand panel), assuming that HE $\gamma$-ray emission has SSC origin. The best fit gives a value of $\gamma_{\min} = 15.05$ for the minimum and $\gamma_{\text{cut}} = 10.99 \times 10^3$ cutoff energies of the electrons.

The magnetic field has a value of $B = 0.19$ G and the radius of the emission region is $R = 5.6 \times 10^{16}$ cm. The fitting result for Doppler factor gives value of $\delta = 11$, consistent with other observations (Liodakis et al. 2017). This means that in the quiescent period the multiwavelength SED is well described by the SSC scenario.

In the flaring periods, EIC is necessary to explain the HE $\gamma$-ray emission (Fig. 6; right-hand panel). The SSC model for the flaring period is below the observed flux, and could not describe flaring fluxes, and the particle cutoff energy ($\gamma_{\text{cut}} = 11.15 \times 10^3$) that characterizes the peak of the energetic parent particle is constrained by the X-ray data. Thus, during the $\gamma$-ray flaring periods, we considered that the blob is located at the outer edge or outside the BLR, where the seed photons originate from the AGN disc and BLR (Donea & Protheroe 2003; Ghisellini, Tavecchio & Ghirlanda 2009). Also, this scenario was accepted to explain the 3-d flaring period of 4C + 28.07 (Das et al. 2021). The typical photon energy from the accretion disc is peaked in the UV, and the IC spectrum could be dominant in HE $\gamma$-ray bands (>100 MeV). The BLR photons peak in optical or near-infrared bands and HE spectrum in this case can extend up to multi-GeV energies and likely up to the Klein–Nishina cut-off. In our calculations, we used the estimated size and Luminosity of the BLR from literature, which are $R_{\text{BLR}} = 4.2 \times 10^{17}$ cm and $L_{\text{BLR}} \sim 2 \times 10^{45}$ erg s$^{-1}$, respectively (Wampler et al. 1984; Cao & Jiang 1999; Shaw et al. 2012; Xiong & Zhang 2014; Isler et al. 2015; Costamante et al. 2018). The optical/UV data from *Swift* are used to estimate the disc luminosity, obtaining $L_{\text{disc}} = 2.7 \times 10^{46}$ erg s$^{-1}$. Generally, the disc luminosity is $L_{\text{disc}} = 10 \times L_{\text{BLR}}$, as suggested in Ghisellini & Tavecchio (2008), which is compatible with the obtained results.

The best fit for EIC model gives a value of $\gamma_{\min} = 28.99$ for the minimum energy of the electrons and $\gamma_{\text{cut}} = 11.15 \times 10^3$ for the cutoff energy. The minimum energy of the electrons is ∼2 times higher compared with SSC scenario; however, the cutoff energy is almost the same (similar) for both cases. A magnetic field of $B = 4.14$ G and $\Gamma_e = 2.25 \pm 0.15$ PL index were required for this EIC model. The Doppler boosting $\delta = 13.7$ factor is a bit higher than the value found for the explanation of the quiescent $\gamma$-ray emission. The total jet energy required ($L_e + L_B \leq 10^{46}$ erg s$^{-1}$) for two models are reasonable, considering the Eddington luminosity for 4C + 28.07 is ∼$2.8 \times 10^{47}$ erg s$^{-1}$. The fitting results for SSC and EIC models are shown in Table 4.

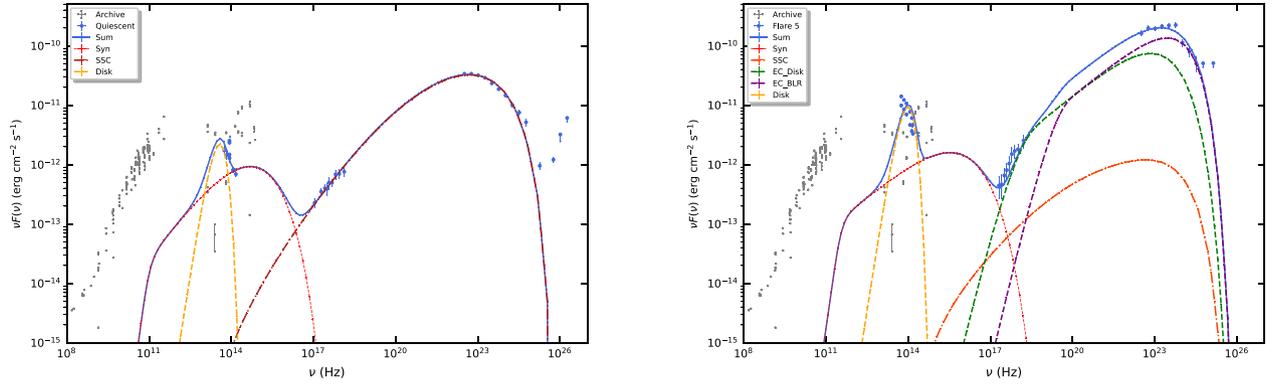

**Figure 6.** Theoretical modelling for the quiescent and flare 5 periods, respectively.

**Table 4.** The modelling results for quiescent and flare 5 periods are presented.

| Period | $\Gamma_e$ | $\gamma_{min}$ | $\gamma_{cut} \times 10^3$ | $R \times 10^{16}$ cm | $B$ (G) | $\delta$ | $U_e$ | $U_B$ | $L_e \times 10^{44}$ erg s$^{-1}$ | $L_B \times 10^{44}$ erg s$^{-1}$ |
|---|---|---|---|---|---|---|---|---|---|---|
| Quiescent(SSC) | 1.8 ± 0.2 | 15.05 ± 1.2 | 10.99 ± 0.31 | 5.6 | 0.19 | 9.86 ± 0.53 | 1.6 | 0.002 | 28.9 | 0.0235 |
| Flare 5 (EIC) | 2.25 ± 0.2 | 28.99 ± 1.2 | 11.15 ± 0.31 | 0.9 | 4.14 | 13.7 ± 0.53 | 0.17 | 0.68 | 1.3 | 5.47 |

## 5 DISCUSSION

4C + 28.07 is one of the brightest FSRQs in the HE $\gamma$-ray band, and multiwavelength studies for this source have been performed by using *Swift-XRT/UVOT* and *Fermi-LAT* data. In the $\gamma$-ray band, the temporal variability studies on weekly time-scales found five flaring intervals, when the flux increased about a factor of ≥5 above the quiescent level (Fig. 1a). These $\gamma$-ray flaring periods are also found in Das et al. (2021), where 3-d time bins flux temporal variations are considered. In this work, in addition the Bayesian Block analysis approach showed rapid $\gamma$-ray variability in daily/hourly scales (Fig. 2).

The corresponding SEDs in the $\gamma$-ray band for the quiescent and each flaring periods are shown in Fig. 4. The complex structure of the SEDs, i.e. early cut-off (<15 GeV) and then hardening at higher energies, could have different explanations. For instance, it can result from $\gamma\gamma$ absorption on He II Lyman photons (Poutanen & Stern 2010). At the same time the $\gamma$-ray spectral hardening can be naturally explained as a $\gamma\gamma$ absorption considering the internal and external narrow-band dense radiation fields of the emitting region (Aharonian, Khangulyan & Costamante 2008; H. E. S. S. Collaboration et al. 2021). Also, the evidence of the cut-off in the $\gamma$-ray spectrum can be connected to the spectral shape and cutoff behaviour of the parent particle distribution (Lefa, Kelner & Aharonian 2012).

During the flaring time periods the source quasi-simultaneously shows activity in the X-ray and UV/optical bands, detected by *Swift-XRT/UVOT*. The multiwavelength SED in Fig. 5 shows that during flaring periods when the source showed activities in the $\gamma$-ray band, the UV/optical and X-ray radiation components are also shifted to higher flux level. This feature is quite promising, since it shows that the physical process taking place in different energy bands are connected, and this can indicate the effective acceleration as revealed by radiation across the electromagnetic spectrum.

To understand the origin of the $\gamma$-ray emission, theoretical modelling has been implemented for the quiescent and flare 5 periods, respectively. The modelling results are depicted in Fig. 6 and the corresponding parameters are shown in Table 4. Within the SSC framework, the $\gamma$-ray emission in low-state level is well explained by the parameters shown in Table 4. The system is out of equipartition ($U_e/U_B \sim 800$), meaning that the radiation site (blob) is particle dominated, full of new and energetic ($\Gamma_e < 2$) *in situ* particles, which are responsible for the $\gamma$-ray emission.

For the flaring period, the EIC model has been applied, assuming external photon contributions from the BLR and disc. Studies have shown that the contribution of an external photon field can effectively influence the EIC radiation for the production of flaring $\gamma$-ray emission (Sikora et al. 1994; Sikora et al. 2009).

Depending on the distance of the emitting region from the SMBH, different scenarios can be discussed, consisting of different physical and external conditions. For example, Costamante et al. (2018) showed that the flaring $\gamma$-ray emission from AGN, which can extend up to 100 GeV, originates from outside of BLR, since the BLR photon field causes $\gamma\gamma$ absorption and therefore the cutoff signature on the $\gamma$-ray spectrum at ∼20 GeV. Here, we assume that the emitting region is located at the outer edge or outside the BLR, where possible photons' contribution can be expected either from disc or BLR. This model is acceptable, considering that for example the rapid (∼10 min) variability of PKS 1222 + 216 is well explained by the outside-of-BLR model (Tavecchio et al. 2011). As a result, the rapid $\gamma$-ray flares can be explained by the energetic and accelerated electrons ($\Gamma_e = 2.25 \pm 0.15$) inside the highly magnetized ($B = 4.14$ G) and compact ($R = 9 \times 10^{15}$ cm) emitting region. These values are similar to the estimated values obtained from previous studies of 4C 28.07 and other blazars (Zhang et al. 2014; Zheng et al. 2017; Anjum, Chen & Gu 2020; Das et al. 2021).

**Table 5.** Comparing derived properties of 4C +28.07 with the sources 3C 279 and M87. References: 4C + 28.07 (Lister et al. 2009; Marshall et al. 2011; Shaw et al. 2012), 3C 279 (de Pater & Perley 1983; Woo & Urry 2002; Zargaryan et al. 2019), and M87 (Biretta, Stern & Harris 1991; Gebhardt & Thomas 2009; Sun et al. 2018).

| Parameters | 4C + 28.07 | 3C 279 | M87 |
| --- | --- | --- | --- |
| Redshift ($z$) | 1.231 | 0.536 | 0.0044 |
| $M_{BH}(10^8\,M_\odot)$ | ∼16 | ∼5 | ∼60 |
| $\Gamma_{jet}$ | 15 | 20 | 5 |
| $L_{\gamma,\,max}$ (erg s$^{-1}$) | $3.6 \times 10^{49}$ | $5.77 \times 10^{48}$ | $10^{42}$ |
| $t_{var}$ | 2 h | 5 min | ∼14 h |
| X-ray jet | Yes | Yes | Yes |
| Knotty structure | Yes | Yes | Yes |

Also, the flaring $\gamma$-rays can result from a clumpy environment inside the blob, which can effectively accelerate and increase the maximum energy of the electrons (Khangulyan et al. 2021). It is quite challenging to explain the short-term variability of these flaring $\gamma$-rays, and further physical explanations are required. Based on the Bayesian Block analysis for the 30-min light curve (see Fig. 2, right-hand panel), the second block size is ∼2 h, which we can consider as a minimum variability time. The emitting region size in this case, assuming $\delta = 10$, will be $R \leq 9 \times 10^{14}$ cm. Taking into account that the BH mass is $1.6 \times 10^9\,M_\odot$ (Shaw et al. 2012), the Schwarzschild radius ($2GM/c^2$) is ∼$4.8 \times 10^{14}$ cm. This means that we deal with ultrafast variability in the scale of the event horizon.

We already discussed the relativistic moving blob inside the jet scenario and, additionally, several other scenarios which can apply to our problem have been proposed to explain the horizon/subhorizon-scale variability in AGN (Aharonian, Barkov & Khangulyan 2017). For instance, the interaction between jets and stars (or clouds) of radius $R_{star} < R_{BH}$ can initiate small (on the scale of the BH gravitation radius) perturbations capable of producing HE $\gamma$-ray emission (Barkov, Aharonian & Bosch-Ramon 2010; Barkov et al. 2012; Khangulyan et al. 2013; del Palacio, Bosch-Ramon & Romero 2019). In this model, it is assumed that the peak of the $\gamma$-ray light curve can be associated with the acceleration of star/blob in the jet.

An alternative scenario is the magnetospheric model when the HE $\gamma$-ray flares can originate in the BH magnetosphere (Beskin, Istomin & Parev 1992; Neronov & Aharonian 2007; Levinson & Rieger 2011; Rieger 2011). However, taking into account that this model is constrained by the flare intrinsic $\gamma$-ray luminosity (e.g. see equation 23 in Aharonian et al. 2017) and considering $L_\gamma \sim 10^{45}$ erg s$^{-1}$ in our case, we can conclude that the rapid flare detected from this source does not have a magnetospheric origin. A more detailed investigation of the alternative scenarios for the particle acceleration in 4C + 28.07 will be given in a future publication.

To highlight the significance of the results obtained for 4C + 28.07, in Table 5 we have compared the results and derived parameters with two other exceptionally well-studied sources, the FSRQ 3C 279 and Radio Galaxy M87. In the radio band, all three sources are intensively observed, bright sources, where extended large-scale jet structure has been detected (de Pater & Perley 1983; Biretta et al. 1991; Lister et al. 2009; EHT MWL Science Working Group et al. 2021). In the X-ray band, M87 and 4C + 28.07 show a complex and extended jet structure, where several knots have been distinguished in *Chandra* observations (Marshall et al. 2011; Sun et al. 2018; Zargaryan, Sahakyan & Harutyunian 2018). At HE $\gamma$-rays, 3C 279 is well known for its rapid flaring on minute scales (5 min) measured by *Fermi-LAT* (Zargaryan et al. 2019). These $\gamma$-ray flares are followed by variability also in the X-ray band observed by *Swift-XRT*. The BH mass for all three sources is $>10^8\,M_\odot$ (Woo & Urry 2002; Gebhardt & Thomas 2009; Shaw et al. 2012). Considering all these similarities between 4C +28.07, 3C 279, and M87, we can see that 4C + 28.07 is a remarkable AGN that should be investigated in more detail.

## 6 CONCLUSION

We report results of the detailed spectral and timing analysis of 4C + 28.07. For this study, we used all available data in the UV/optical, X-ray, and $\gamma$-ray bands, detected by *Fermi-LAT*, *Swift-XRT/UVOT*, and telescopes. Five flaring periods have been detected in the $\gamma$-ray band. The brightest and most rapid (∼30 min) flare (flare 5) was observed on 2018 October 12, when the $\gamma$-ray flux reached $(6.7 \pm 0.81) \times 10^{-6}$ photon cm$^{-2}$ s$^{-1}$, about 31 × higher than the mean flux over 12.3 yr, with detection significance of $\sigma = 6.1$. Next, the spectrum of the ∼12.3 yr of $\gamma$-ray data extends up to ∼316 GeV, displaying complex structure by having early cut-off at ∼14 GeV, and then hardening beyond ∼60 GeV. Also, similar spectral features are apparent in the SEDs of flares. The total ∼12.3 yr of detected $\gamma$-ray emission was well described by the SSC scenario. Regarding the explanation of the rapid flares, the EIC scenario has been discussed by assuming external photon contributions from the disc and the BLR. The estimated $\gamma$-ray emitting region size ($R \leq 9 \times 10^{14}$ cm) is close to the BH event horizon, which makes it challenging to explain the origin of rapid flares and brings 'subhorizon' AGNs ultrafast variability scenarios into discussion. Giving consideration to the remarkable $\gamma$-ray emission detected from 4C 28.07, we made a comparison with 3C 279 and M87 to highlight 4C + 28.07 as another excellent laboratory for further investigation of AGN physics.

## ACKNOWLEDGEMENTS

DZ acknowledges funding from an Irish Research Council (IRC) Starting Laureate Award (IRCLA\2017\83). JM acknowledges support from a Royal Society-Science Foundation Ireland University Research Fellowship (14/RS-URF/3219 and 20/RS-URF-R/3712) and an Irish Research Council (IRC) Starting Laureate Award (IRCLA\2017\83). We thank the referee for their comments which helped to improve the manuscript. We acknowledge the use of public data from the *Fermi-LAT* and *Swift* data archive.

## DATA AVAILABILITY

All of the data underlying this article are publicly available and can be downloaded from the *Fermi-LAT* and *Swift* data servers.

This paper has been typeset from a T<sub>E</sub>X/L<sup>A</sup>T<sub>E</sub>X file prepared by the author.